\documentclass[useAMS]{mn2e}
\usepackage{psfig}

\usepackage{times}

\title[Rapid variability in TeV blazars] 
{Rapid variability in TeV blazars: the case of PKS 2155--304 }

\author[Ghisellini \& Tavecchio]
{G. Ghisellini$^{1}$\thanks{E--mail: gabriele.ghisellini@brera.inaf.it}, 
F. Tavecchio$^{1}$ \\
$^{1}$INAF -- Osservatorio Astronomico di Brera, Via E. Bianchi 46, I-23807
Merate, Italy
}

\begin{document}

% \date{Accepted 1988 December 15. Received 1988 December 14; 
% in original form 1988 October 11}

%\pagerange{\pageref{firstpage}--\pageref{lastpage}} \pubyear{2007}

\maketitle

\begin{abstract}
Recent Cherenkov observations of BL Lac objects showed 
that the TeV flux of 
PKS 2155--304 changed by a factor 2
in just 3--5 minutes.
This fast variability can be accounted for if the emitting region is
moving with a bulk Lorentz factor $\Gamma\sim 50$ and a similar 
relativistic Doppler factor.
If this $\Gamma$ is adopted, several models can fit the data,
but, irrespective of the chosen model, the jet is matter dominated. 
The Doppler factor  
requires 
viewing angles of the order of $1^\circ$ or less: 
if the entire jet is as narrow as this, then
we have problems with current unification schemes.
This suggests that there are small active regions, inside a larger jet, 
moving faster than the rest of the plasma, occasionally
pointing at us.
Coordinated X--ray/TeV variability can discriminate between 
the different scenarios.
\end{abstract}

\section{Introduction}

The blazars PKS 2155--304 (Aharonian et al. 2007) and Mkn 501  
(Albert et al. 2007) 
showed variations on $t_{\rm var}=$3--5 minutes of their TeV flux.
This ultrafast variability calls for a revision of our current
ideas about the properties of the emitting region.
Particularly challenging is the case of PKS 2155--304, because the
variations occurred during an overall very active state of the source
with an observed TeV luminosity of $\sim 10^{47}$ erg s$^{-1}$.
The usual way to infer the size from the variability timescale
is $R<ct_{\rm var}\delta$ where $\delta$ is the Doppler factor.
Begelman, Fabian \& Rees (2008; hereafter B08) 
pointed out that such small $t_{\rm var}$ cannot be any longer indicative of 
the size of the black hole (as instead are in the 
``internal shock scenario" (Sikora et al 1994, Ghisellini 1999, 
Spada et al. 2001, Guetta et al. 2004).
The mass of the central black hole of PKS 2155--304 is 
uncertain (of the order of $10^9$ $M_\odot$; Aharonian et al. 2007)
but even a (prudently small) mass of $M=10^8$ solar masses
corresponds to a light travel time across the Schwarzschild 
radius of $10^3$ s.
To avoid to have a too compact source, with
the accompanying problem of $\gamma$--ray absorption through the 
$\gamma$--$\gamma$ $\to e^\pm$ process (see e.g. Dondi \& Ghisellini 1995), 
the bulk Lorentz factor of the emitting
region (hence $\delta$) must have a value close to 50.
% (close to the ones of Gamma Ray Bursts). 
This requirement is not unprecedented, since 
also the large separation, in frequency, of the two broad peaks 
characterising the spectral
energy distribution (SED) requires $\delta>30$
in single--zone synchrotron self Compton (SSC) models 
(see e.g. Konopelko et al. 2003).
Smaller $\delta$ and $\Gamma$ factors are however possible in
alternative models, where two or more emission
regions have different $\Gamma$, as in the decelerating jet model 
of Georganopoulos \& Kazanas (2003) and in the ``spine--layer" model proposed 
by Ghisellini, Tavecchio \& Chiaberge (2005; hereafter G05).
In both models contiguous emitting regions of the jet are moving
with different $\Gamma$, and one part can see the emission of other 
relativistically boosted, thus enhancing the radiation energy density
and the consequent inverse Compton emission.
In this way one can relax the requirement of having $\Gamma$--factors
of the order of 30 or more, requiring a more standard value of around 15.
The observed rapid variability, instead, cannot be ``cured" by
invoking a structured jet emission region, and is therefore
more demanding.

The implied large $\Gamma$ makes any external photon source strongly
boosted in the comoving frame, favouring the ``external Compton" (EC)
process, in which the energy density of the seed photons produced
outside the jet is larger than the energy density of the synchrotron
radiation produced locally.  
This argument leads B08 to suggest that the EC process is the main 
radiation mechanism.  
They also argue that,
since to produce TeV photons one needs highly energetic electrons
(with random Lorentz factors $\gamma \sim 10^6/\Gamma$), the jet could be
particle starved (i.e. one needs fewer electrons to produce the
radiation we see, if they are at high energies).  Therefore the jet
should be magnetically dominated, i.e. the power carried by the jet
should be mainly in the form of the Poynting flux associated to
magnetic fields.

In this paper we investigate if the two claims: 1) EC favoured with
respect to the SSC model, and 2) the jet is magnetically dominated
(made by B08 on the basis of rather general arguments), are confirmed
by detailed modelling of the SED of PKS 2155--304 with the SSC and the EC models.
In particular, we consider two constraints not considered in B08:
i) there is a limit on the amount of the seed external luminosity,
posed by the observed flux, and ii) the predicted synchrotron flux 
must reproduce, or be smaller, than the observed optical/X--ray flux.
We also present a new scenario, analogous to the spine/layer model,
in which a compact and active region is moving within a larger jet,
with a bulk Lorentz factor larger then the one of the surrounding plasma. 
We call this version of the spine/layer model the ``needle/jet'' model.

\section{Data and models}

The SED of PKS 2155--304 is reported in Figs. \ref{2155},
\ref{2155nj}.  The shown TeV data correspond to
the flare of July 28, 2006 (Aharonian et al. 2007).
Red symbols are 
the de--absorbed fluxes, obtained with the ``low SFR'' model of
Kneiske et al. (2004), consistent with the recent measures indicating
a low value of the IR intergalactic background (Aharonian et al. 2006). 
Unfortunately there are no observations in other
bands exactly simultaneous to the TeV flare of July 28, 2006.
Therefore we do not know if the X--ray flux was also varying as
rapidly, together with the TeV flux.  On the other hand, PKS 2155--304
was observed on July 29--31 and several time on August, 2006 by the
{\it Swift} satellite (see Foschini et al. 2007, hereafter F07).  The
X--ray and the optical data of July 30 and August 2 are shown in both
Fig. \ref{2155} and Fig. \ref{2155nj} (see F07 for details of the
analysis).  Another observation by {\it Chandra} (partially
overlapping with the first one by {\it Swift}) is still unpublished.
PKS 2155--304 is one of the best studied extragalactic X--ray sources,
and it never showed large changes of its X--ray spectral shape
(unlike Mkn 501, whose
synchrotron peak frequency changed by more than a factor 100).  We
therefore assume that the X--ray flux and slope observed on July 30
were representative of the state during the TeV flare.  The other
data, coming from the literature, are neither simultaneous nor close
in time with the TeV data, but can give a rough representation of the
entire SED, since at frequencies smaller than the synchrotron peak the
flux variability is much reduced.

We will use three emission models to interpret the SED of PKS
2155--304 and in particular the rapidly variable and active TeV state
observed on July 28, 2006.  In all cases we will assume $\Gamma=50$
and a source size $\sim 3 \times 10^{14}$ cm, in order to have 
$t_{\rm var}\sim 200$ s.  
The magnetic field $B$ is tangled and homogeneous.
For simplicity, the inverse Compton emission is calculated assuming a
truncated scattering cross section, equal to the Thomson one if
$\gamma h\nu^\prime /m_e c^2 <3/4$, and 0 otherwise ($\gamma m_e c^2$
is the electron energy and $\nu^\prime$ is the photon frequency as
seen in the comoving frame of the source). This mimics the real
decline, with energy, of the Klein--Nishina cross section only
approximately, sufficient if we are not interested to precisely
describe the TeV spectrum, as in our case, but poor if we want, for
instance, reconstruct the IR background through the detailed modelling
of the TeV blazar spectrum.

{\bf SSC ---} The first model is a simple one--zone pure SSC model.
As shown by Tavecchio, Maraschi \& Ghisellini (1998), in this case the
number of parameters characterising the model is equal to the number
of observables, and there is no freedom if the frequencies and flux
levels of the two peaks of the SED are known (and the variability
timescale, as in our case, is given).

{\bf EC ---} The second model is an EC model. 
To fix the ideas we
assume that the dominant source of seed photons is the jet beyond the
TeV region. 
Indeed there are evidences supporting the view that
the jet, soon after the location of the TeV emission, strongly
decelerates.
In fact, at the pc scale, VLBI data show knots
moving with an apparent speed of $\beta_{\rm app} = 4.4\pm2.9$ (Piner
\& Edwards 2004, assuming the Hubble constant $H_0=71$ km s$^{-1}$
Mpc$^{-1}$). The model parameters are chosen in order to maximise the
contribution of the seed photons for the scattering process and the
value of the magnetic field in the TeV emitting region.  In particular
we assume that the emission peaks in the mm/submm part of the
spectrum, close to where the self--absorption frequency is expected.

{\bf Needle/Jet ---} The third model assumes that the ``persistent''
(e.g. varying on $\sim$hours timescales) emission is produced by a jet
whose plasma moves with the standard $\Gamma=15$ and that the
variable flux originates in a small portion of the jet (a ``needle''),
moving at $\Gamma=50$.  The latter, being immersed in the radiation
field produced by the jet, sees an enhanced radiation energy density.

In the following subsections we briefly summarise the set up of
the three different models.

\subsection{SSC and External Compton models}

These models are described fully in Celotti \& Ghisellini (2008;
hereafter CG08).
The observed radiation is postulated to originate in a single zone of
the jet, described as a cylinder of cross sectional radius $R$ and 
thickness (as seen in the comoving frame) $\Delta R'= R$.

The relativistic particles are assumed to be injected throughout the
emitting volume for a finite time $t^\prime_{\rm inj}=\Delta R'/c$.
The observed (flaring) spectrum is obtained by considering
the particle distribution at the end of the injection, at
$t=t^\prime_{\rm inj}$, when the emitted luminosity is maximised. 

As the injection lasts for a finite timescale, not all the 
particles have time to cool in the time $t^\prime_{\rm inj}$,
but only those particles with energies greater than the cooling
energy $\gamma_{\rm c}$.
The particle distribution $N(\gamma)$ can be described as a broken
power--law with a slope equal to the injection slope below $\gamma_{\rm c}$ 
and steeper above it.
% The  $N(\gamma)$ distribution
% corresponds 
We assume a particle injection function extending from $\gamma=1$ to
$\gamma_{\rm max}$, with a broken power--law shape with slopes $\propto
\gamma^{-1}$ and $\propto \gamma^{-s}$ below and above  $\gamma_{\rm inj}$. 
The resulting shape of $N(\gamma)$ depends on the injected 
distribution and on the cooling time with respect to $t_{\rm inj}$.

If the cooling time is shorter
than $t_{\rm inj}$ for particle energies $\gamma<\gamma_{\rm inj}$  
(fast cooling regime) the resulting $N(\gamma)$ is 
\begin{eqnarray}
N(\gamma) & \propto & \gamma^{-(s+1)};  
\qquad  \propto \gamma^{-2}; 
\qquad \quad \propto \gamma^{-1}; 
\nonumber\\
\gamma &>& \gamma_{\rm inj};
\qquad \gamma_{\rm c}   <  \gamma < \gamma_{\rm inj};
\qquad \gamma  <  \gamma_{\rm c}
\end{eqnarray}
This is the case of our EC model.
In the opposite case, only those electrons with $\gamma>\gamma_{\rm c}> \gamma_{\rm inj}$
can cool in $t_{\rm inj}$ (slow cooling regime), and we have
\begin{eqnarray}
N(\gamma) &\propto& \gamma^{-(s+1)}; 
\qquad \propto \gamma^{-s};
\qquad \quad \,\, \propto \gamma^{-1}  \nonumber\\
\gamma &>& \gamma_{\rm c};
\qquad \quad \gamma_{\rm inj} <\gamma < \gamma_{\rm c};
\qquad \gamma < \gamma_{\rm inj}.
\end{eqnarray}
This is the case of our SSC model.
For the EC model, the external radiation of luminosity $L_{\rm ext}$
is assumed to be produced at a distance $R_{\rm ext}$ from the black hole.
Since we assume that this source is stationary with the black hole, in the comoving
frame of the TeV emitting source it is seen strongly boosted
and blue--shifted.
The external emission is assumed to be peaked at the frequency $\nu_{\rm ext}$ 
and distributed as a black body (for simplicity).
In the cases presented here, the random Lorentz factor of
the electrons emitting at the peaks of the SED, $\gamma_{\rm peak}$, always
coincides with $\gamma_{\rm inj}$.

\subsection{Needle/Jet model}

The model is similar to the spine/layer models described in G05 (see
also Tavecchio \& Ghisellini 2008).
An analogous idea was also proposed in the Gamma Ry Burst field
to explain the fast variability of the prompt emission (Lyutikov 2006).
As in G05, we have a compact
region moving faster then the rest of the jet.  In G05 this was the
entire ``spine" of the jet, while the slower material was supposed to
be a (thin) boundary layer surrounding it.  Both components, in G05,
had a strong radiative interplay since one sees the radiation produced
by the other relativistically boosted.  Here, instead, the compact and
fast region is thought as a ``needle" moving throughout a larger jet.
The velocity vector of the needle could even be not perfectly aligned
with that of the larger jet (oblique shocks?), but for simplicity we will
assume that it is.  It illuminates the larger jet, but due to its
larger $\Gamma$--factor the fraction of the illuminated jet is small.
Consequently, we can neglect the contribution of the needle to the
radiation energy density seen in the jet frame.  The reverse of course
is not true: the needle travels in a dense radiation field, provided
by the jet emission.  The idea is that the entire jet is responsible
for the persistent emission of PKS 2155--304, and the ``needle" is
responsible for the very rapid variability.

Both the jet and the needle are filled by a tangled magnetic field and by 
relativistic electrons assumed to follow a
smoothed broken power--law distribution extending from $ \gamma_0>1$
to $\gamma_{\rm max}$ with slopes $\gamma^{-n_1}$, $\gamma^{-n_2}$ below and above the
break at $\gamma_{\rm peak}$.  
The normalisation of this distribution is
calculated assuming that the system produces an assumed synchrotron
luminosity $L^\prime_{\rm syn}$ (as measured in the local frame), which is an
input parameter of the model.  

In the comoving frame of the needle the photons produced by the jet
are not isotropic, but are seen aberrated.  Most of them are coming
from almost a single direction (opposite to the relative velocity
vector).  In this case the produced Compton flux is anisotropic even
in the comoving frame, with more power emitted along the opposite
direction of the incoming photons (i.e. for head--on scatterings).
Consider also that this applies only to the Compton scatterings with
photons coming from the jet, while the synchrotron and the SSC
emissions are isotropic in the comoving frame.  The EC radiation
pattern of the needle will be more concentrated along the jet axis (in
the forward direction) with respect to its synchrotron and SSC
emission.  The corresponding transformations are fully described in
G05 and Tavecchio \& Ghisellini (2008).

\subsection{Jet power}

Our modelling allows to derive all the quantities needed to calculate
the power carried by the jet in different forms: bulk motion of cold
protons and of energetic electrons, magnetic field and radiation.  To
each of these components there is a corresponding energy density
$U^\prime_i$ as measured in the comoving frame of the TeV emitting
source.  For instance, for the electron component,
$U^\prime_i=U^\prime_e=\int N(\gamma) \gamma m_e c^2 d\gamma$.  We
consider only those electrons needed to explain the emission.  Since
other (cold) electrons can be present, the derived power is a lower
limit.  For the proton component, we assume one proton per emitting
electron.  If the emitting leptons include electron--positron pairs,
then the derived proton energy density and power are upper limits.
The power is calculated through the flux across a surface
perpendicular to the jet axis: $L_i \equiv \pi R^2 \Gamma^2 c\,
U^\prime_i$.

%--------------------------------------------------
\begin{figure}
\vskip -0.5 cm 
\centerline{
\psfig{figure=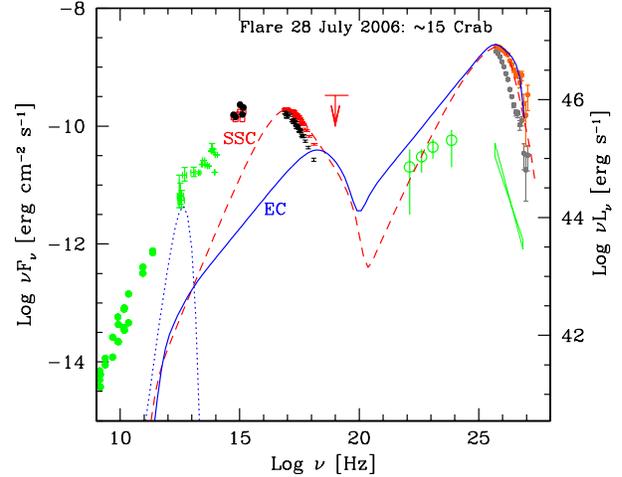,width=9cm,height=9cm}}
\vskip -1.7 cm 
\caption{The SED of PKS 2155--304. Observed TeV data from
H.E.S.S. (grey) correspond to the flare of 28 July 2006 (Aharonian et
al. 2007). The red points report the TeV spectrum corrected for the
extragalactic absorption (see text for details). X--ray and optical
data are not strictly simultaneous to the TeV ones, but corresponds to
2 and 4 days later (see F07).  
Green symbols are archival data.
We have tried to model the SED with an
SSC (dashed line) and EC (solid line) assuming the parameters listed
in Table 1. 
The dotted line corresponds to the distribution of seed photons assumed 
for the EC model.
It has been chosen in order to maximise the high energy output of the EC 
emission, without overproducing the far IR observed flux.
}
\label{2155}
\end{figure}
\begin{figure}
\vskip -0.5 cm 
\centerline{
\psfig{figure=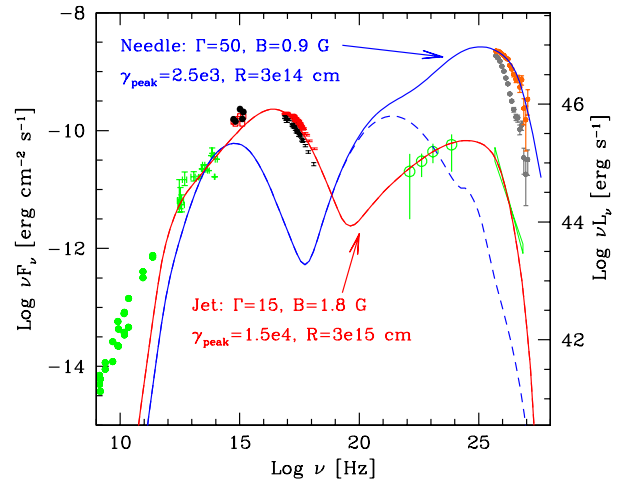,width=9cm,height=9cm}}
\vskip -1.7 cm 
\caption{
As in Fig \ref{2155}, but the model here corresponds to a needle/jet scenario.
The dashed line corresponds to the flux produced by the needle if we neglect
the radiation energy density produced by the layer.
The radiation energy density seen by the jet
due to the needle emission is assumed to be negligible (see text).}
\label{2155nj}
\end{figure}
%--------------------------------------------------

\section{Results}

Fig. \ref{2155} shows the TeV data corresponding to the flare
(observed and de--absorbed), 
% according to a low value of the IR background), 
together to X--ray and optical data taken a few days later (F07). 
The shown models are the pure SSC and the EC model, the
adopted parameters are listed in Table 1.  
%Both models can reproduce the high energy data, but the EC model
%cannot contribute much to the X--ray flux, which can instead be
%reproduced by the SSC model.
The external source of seed photons for the EC model, represented by
the dotted line in Fig. \ref{2155}, is assumed to be stationary with
the black hole.  This is likely to be not realistic, but in line with
the aim to find a limit to the maximum amount of possible seed
photons, in the submm band, produced externally to the TeV emitting
region.

The SSC model can satisfactorily reproduce the entire SED. For the EC
model, we have choosen a solution maximizing the value of the magnetic
field in the emitting region, to test the possibility to have a
magnetically dominated jet. 
Large magnetic fields, coupled with large electron energies
(necessary to emit TeV photons),
yield a synchrotron peak at frequencies larger than observed. 
This, in turn, requires that the synchrotron flux in the EC model cannot
contribute much to the low frequency peak of the SED, requiring
another component producing the radiation below the hard X--ray band.
%
% ---------------------------------------------
% \setcounter{table}{0}
\begin{table}
\begin{tabular}{lllllll}
\hline 
                   &SSC    &EC       &Needle    &Jet     &F07    &Units  \\
\hline
$\Gamma$           &50     &50       &50        &15      &30     &        \\     
$\theta$           &1      &1        &1         &1       &1.7    &degree  \\  
$R$                &3.6e14 &3.2e14   &3e14      &3e14    &5e15   &cm      \\     
$\Delta R^\prime$  &3.6e14 &3.2e14   &3e14      &3e15    &5e15   &        \\     
$B$                &0.58   &3        &0.9       &1       &0.27   &Gauss   \\  
$L^\prime$         &5e40   &3.5e40   &---       &---     &1.1e42 &erg s$^{-1}$   \\
$L^\prime_{\rm syn}$ &---  &---      &3.7e39    &8e40    &---    &erg s$^{-1}$   \\
$L_{\rm ext}$      &---    &2e44     &---       &---     &---    &erg s$^{-1}$   \\  
$R_{\rm ext}$      &---    &3e17     &---       &---     &---    &cm   \\     
$\nu_{\rm ext}$    &---    &4e12     &---       &---     &---    &Hz   \\     
$\gamma_{\rm max}$ &6e5    &2e5      &3e5       &7e5     &1.8e5  &     \\  
$\gamma_{\rm peak}$&3e4    &5e4      &9e3       &1.5e4   &1.5e4  &     \\ 
$\gamma_0$         &1      &1        &4e2       &1e2     &1      &     \\ 
% $\gamma_{\rm cool}$&1.7e4  &149       &        &       &       &   \\   
$n_1$              &---    &---      &2         &2       &---    &     \\   
$n_2$              &4.2    &3.6      &4.7       &4.3     &3.5    &     \\  
\hline
$L_{\rm p}$        &3.4e43 &9.9e42   &3.4e44    &4.4e43  &6.1e44 &erg s$^{-1}$   \\  
$L_{\rm e}$        &6.0e43 &9.1e41   &1.7e44    &1.2e43  &1.9e44 &erg s$^{-1}$   \\  
$L_{\rm B}$        &4.1e41 &8.6e42   &6.8e41    &2.4e43  &2.6e43 &erg s$^{-1}$   \\  
$L_{\rm r}$        &8.0e43 &8.8e43   &1.7e43    &2.5e43  &2.7e44 &erg s$^{-1}$   \\  
\hline
\end{tabular}
\caption{
Parameters for the models used to explain the SED of PKS 2155--304
during the energetic and rapid flare of July 28, 2007.
The last column (F07) reports the values of the SSC model
used to fit the SED
of PKS 2155--304 on the day after (i.e. July 29, 2006,
when we have the simultaneous TeV and X--ray data, see F07). 
For all models but the last two, the Doppler factor is $\delta=56.8$
(it is $\delta=28$ for the ``jet" and $\delta=33.5$ for F07).
}
\end{table}

Fig. \ref{2155nj} shows the same data and the results of the 
needle/jet model.  
The dashed line shows the emission of the needle, neglecting 
the contribution of the radiation produced by the jet.  
The red solid line is the emission for the larger jet.  
The needle is assumed to move with $\Gamma=50$ while the
jet has $\Gamma=15$.  Note the strong enhancement of the inverse
Compton flux of the needle due to the contribution of the jet
radiation energy density.

Both the X--ray and the TeV flux are reproduced by the model,
but the TeV flux is produced by the needle, and the X--ray flux is
produced by the jet.
The optical--IR flux receives comparable contributions from both.
At these frequencies, the SSC and EC model have a large deficit.
For the SSC model, this is due to the flatness of the 
electron distribution function: 
since the cooling time for IR emitting electrons is long, the
emitting particle distribution retains the injection slope 
($\propto \gamma^{-1}$) after a time $\sim R/c$.
The deficit at IR frequencies of the EC model comes instead from
the requirement of not overproducing the X--ray flux.
Note in fact that the slope of the EC model before the synchrotron
peak is softer than in the SSC model, since in this case even low
energy electrons do cool in $R/c$, making the particle distribution
$\propto \gamma^{-2}$ down to $\gamma \sim 150$. 

For all models we have checked that the $\gamma$--$\gamma \to e^\pm$ process
is not important, taking also into account (when appropriate)
the target photons produced externally to the TeV emitting region.

\begin{figure}
\vskip -0.8 cm 
\centerline{
\psfig{figure=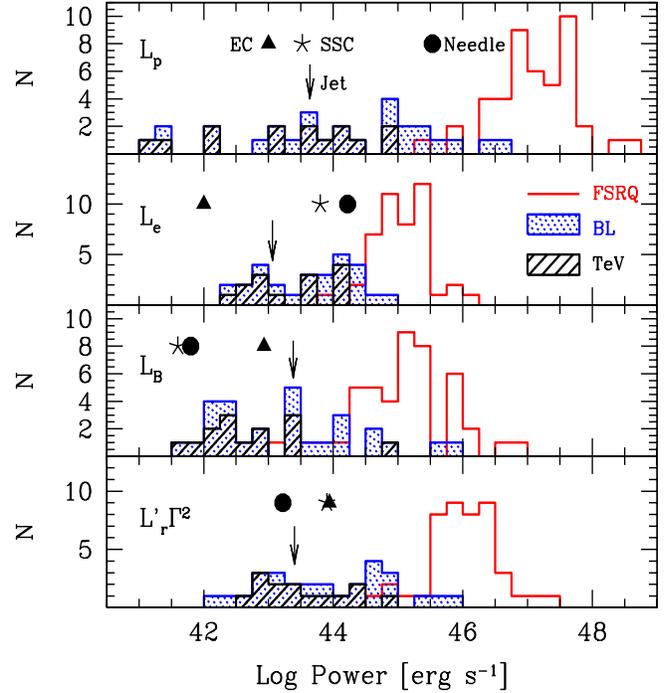,width=10cm,height=11cm}}
\vskip -0.9 cm 
\caption{
The power of blazar jets in bulk motion of protons
($L_{\rm p}$, assuming one proton per emitting electrons),  
relativistic electrons ($L_{\rm e}$), magnetic field ($L_{\rm B}$) and
in radiation $L_{\rm r}$.
The star, triangle, circle and arrow correspond to the values of 2155--304
during the TeV flare of 28 July 2006, whose SED has been fitted
with $\Gamma=50$ and with the SSC, the EC and the needle/jet models 
(see Fig. \ref{2155} and Fig. \ref{2155nj}).
These values are compared with the ones derived for a sample 
of blazars by CG08.
}
\label{isto}
\end{figure}

In Fig. \ref{isto} we show the jet powers of PKS 2155--304 according
to the three models, and compare these values with the ones derived
for the sample of $\gamma$--ray emitting blazars studied by CG08.
Note that $L_{\rm B}$ is not dominant even in the EC model, despite
the fact that the enhanced radiation energy density (due to the strong
boosting) allows a larger magnetic field with respect to the SSC model
($B=3$ G vs 0.58 G, respectively).  
Note also that the power $L_{\rm p}+L_{\rm e}$  barely corresponds to 
$L_{\rm r}$ in the SSC case, and is smaller in the EC model.  
The needle of the needle/jet model has the largest
$L_{\rm p}$, and a Poynting flux similar to the one of the SSC model.
As expected, the values of the jet of the needle/jet model are
comparable to the other TeV blazars considered in CG08.

\begin{figure}
\vskip -0.5 cm 
% \centerline{
\hskip -2 cm
\psfig{figure=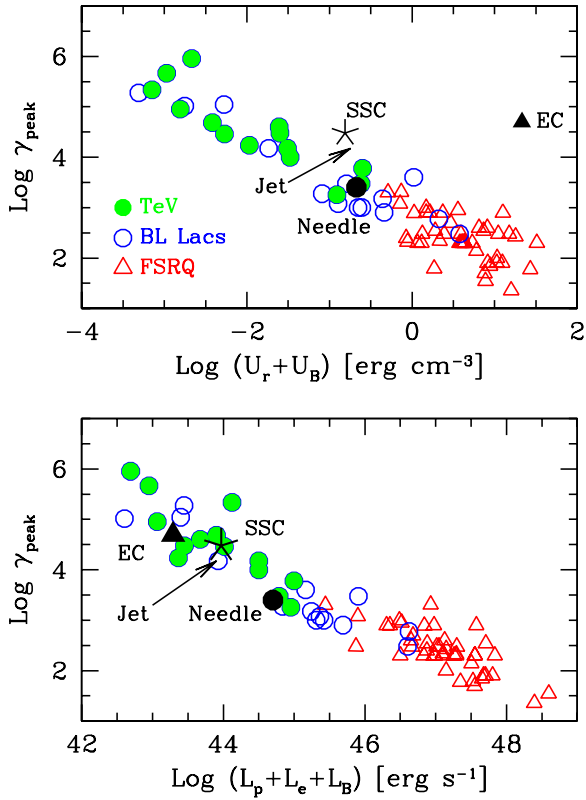,width=12cm,height=12cm}% }
\vskip -0.8 cm 
\caption{
The random Lorentz factor, $\gamma_{\rm peak}$, of the electrons
emitting at the peaks of the SED as a function of the comoving energy density
(magnetic plus radiative; top panel) and the total jet power (bottom panel).
The star, triangle, circle and arrow correspond to the values of PKS 2155--304
during the TeV flare of 28 July 2006, whose SED has been fitted
with $\Gamma=50$ and with the SSC, the EC and the needle/jet models 
(see Fig. \ref{2155} and Fig. \ref{2155nj}).
These values are compared with the values of a sample of blazars discussed
in CG08.
 }
\label{gpeak}
\end{figure}

In Fig. \ref{gpeak} we show the value $\gamma_{\rm peak}$ of PKS
2155--304 vs the comoving energy densities and jet power.  The SSC and
the needle/jet model give values of $\gamma_{\rm peak}$ and $U^\prime$
in reasonable agreement with the general trend observed for other TeV
blazars, fitted with a smaller $\Gamma$.  The values for the EC and
spine--layer models, instead, stand clearly apart.  The correlation
between $\gamma_{\rm peak}$ and the jet power, $\gamma_{\rm
peak}\propto L_{\rm j}^{-3/4}$, is well followed by all three models.

%
%--------------------------------------------------
%--------------------------------------------------

%--------------------------------------------------
%--------------------------------------------------

\section{Discussion}

We have modelled the SED of PKS 2155--304 during the huge TeV flare of
July 28, 2006 with three models, assuming always a bulk Lorentz factor
$\Gamma=50$, a size of the TeV emitting region $R\sim 3\times 10^{14}$
cm, and a viewing angle of $1^\circ$, to account for the observed
short variability timescale of the TeV flux.  As mentioned in the
Introduction, our aim was to check whether (as concluded by B08) in
this case 
1) the large $\Gamma$ favours the dominance of seed photons produced
externally to the jet to produce the TeV flux, and 
2) the jet (or the portion of the jet) producing the
variable TeV emission could be Poynting flux dominated.

The answer to 2) is negative since the Poynting flux
carried by the jet is never dominating, even assuming the most
favourable conditions. 
It is true that, in the EC case, the larger the external radiation energy density
the larger the allowed magnetic field (and thus $L_{\rm B}$),
but the limit posed by the SED itself to the amount of the luminosity
produced externally to the TeV emitting region is also limiting the maximum
value of the magnetic field.  
% Both the SSC and 
% Also the spine--layer model can reproduce the data very well,
% but it has almost twice the number of free parameters 
% than the SSC model (one set for the spine, and another set for the layer),
% so the good match with the data is not surprising.
% However, even in this case, the jet power is matter dominated
% if we assume that there is one proton per emitting electron,
% and nearly in ``equipartition" (i.e. $L_{\rm B}\sim L_{\rm e}\sim L_{\rm r}$)
% if there is a significant number of electron--positron pairs.

The answer to 1) (is the external radiation important?) is more complex.
In our fits the SSC model reproduces the data better than the EC model. 
However, this is due to the fact that we wanted to fit the high energy 
data and, {\it at the same time}, maximise the magnetic field. 
Allowing a {\it smaller} magnetic field we could obtain a
better agreement with the SED.  
In fact the best fitting model is the needle/jet one,
that can be thought as a kind of EC model, where the
``external" radiation is the radiation produced by the jet itself, but
externally to the needle.

In the EC case, the power carried by the jet in emitting electrons is
smaller than what carried in radiation.  This apparent paradox (how
can the radiation have more power than the electrons producing it?) is
solved recalling that in this EC model the cooling time of the
electron becomes short, with the need of refreshing the electron
distribution with the injection of new particles.  Therefore the flow
of relativistic electrons across a given jet cross sectional area
underestimates the number (and the energy) of the electrons having
contributed to the radiation flux. 
The derived $L_{\rm e}$ is therefore a lower limit.

Concerning the ``blazar sequence", in the form of the relation between
$\gamma_{\rm peak}$ and the energy density $U^\prime$ (magnetic plus
radiative) seen in the comoving frame, it is well obeyed in the SSC
and needle/jet case, not so for the EC model.  Intriguingly, the
relation between $\gamma_{\rm peak}$ and the total jet power is
instead followed by all the three models.

The SSC and EC models here discussed cannot describe the ``quiescent"
or ``persistent" state of blazars,  (but in any case variable on 
timescale of the order of $10^4$ s) since the combination of jet 
narrowness and the required small viewing angle cannot be reconciled 
with current unification schemes.  
One possibility is that PKS 2155--304 is unique, and the fast TeV 
variability will not be a general property of TeV blazars.  
This is unlikely, for two reasons: the first is that
already another TeV blazars showed a very small $t_{\rm var}$
(i.e. Mkn 501), the other reason is that such a fast variability was
observed as soon as the Cherenkov telescopes had the required
sensitivity to detect it.

It may be dangerous to infer, from the ultrafast TeV variability,
the general properties of a jet that can be
discontinuous, either in space or in time, or both.
In other words: episodes of very fast variability can be
produced by ``needles" of matter moving faster than average,
oriented in different directions but contained in a 
wider cone, and occasionally pointing at the observer. 
Or the emitting regions are shells that corresponds to an
intermittent activity of the central engine, having on average
a bulk Lorentz factor that is quite large but not extreme,
and occasionally going much faster (and possibly contained in a narrower cone).
In the first case a ``needle" would move through a region
already filled with radiation (produced by the rest of the jet),
and this would resemble the needle/jet model here discussed,
in the second case it is likely that the emitting region would be
rather external--photon--starved, and the SSC model could better describe
the corresponding SED.
How to distinguish the two cases?
A crucial point is the UV--X--ray flux, if observed exactly 
simultaneously with the TeV band. 
If there is no sign of simultaneous fast variability, then 
the ``needle" model is to be preferred, since the UV--X--rays
could be produced by the entire jet, while the TeV flux can come
from the needle.
If instead both the UV--X--ray and the TeV fluxes vary together 
and quickly, then the SSC model is favoured.

% \section*{Acknowledgments}
% We thank 

\end{document}